\documentclass{aastex631}
\usepackage{longtable}
\usepackage{comment}

\usepackage{amsmath, physics}
\shorttitle{A Parkes ``Murriyang'' Search for Pulsars and Transients in the LMC}
\shortauthors{Hisano et al.} 

\begin{document}

\title{A Parkes ``Murriyang" Search for Pulsars and Fast Transients in the Large Magellanic Cloud}

\author[0000-0002-7700-3379]{Shinnosuke Hisano}
\affiliation{Kumamoto University, Graduate School of Science and Technology, Kumamoto, 860-8555, Japan}
\email{204d7152@st.kumamoto-u.ac.jp}

\author[0000-0002-2578-0360]{Fronefield Crawford}
\affiliation{Department of Physics and Astronomy, Franklin \& Marshall College, P.{}O.~Box~3003, Lancaster, \hbox{PA}, 17604 USA}

\author{Victoria Bonidie}
\affiliation{Department of Physics and Astronomy, Franklin \& Marshall College, P.{}O.~Box~3003, Lancaster, \hbox{PA}, 17604 USA}

\author{Md F. Alam} 
\affiliation{Department of Physics and Astronomy, Franklin \& Marshall College, P.{}O.~Box~3003, Lancaster, \hbox{PA}, 17604 USA}

\author[0000-0002-3034-5769]{Keitaro Takahashi}
\affiliation{Kumamoto University, Graduate School of Science and Technology, Kumamoto, 860-8555, Japan}
\affiliation{Kumamoto University, International Research Organization for Advanced Science and Technology, 2-39-1 Kurokami, Chuo-ku, Kumamoto 860-8555, Japan}
\affiliation{National Astronomical Observatory of Japan, 2-21-1 Osawa, Mitaka, Tokyo 181-8588, Japan}

\author[0000-0003-1301-966X]{Duncan R. Lorimer}
\affiliation{Department of Physics and Astronomy, West Virginia University, P.O. Box 6315, Morgantown, WV 26506, USA}
\affiliation{Center for Gravitational Waves and Cosmology, West Virginia University, Chestnut Ridge Research Building, Morgantown, WV 26505, USA}

\author[0000-0002-3017-092X]{Josh P. Ridley}
\affiliation{School of Engineering, Murray State University, Murray, KY~42071, USA}

\author[0000-0001-7697-7422]{Maura M. McLaughlin}
\affiliation{Department of Physics and Astronomy, West Virginia University, P.O. Box 6315, Morgantown, WV 26506, USA}
\affiliation{Center for Gravitational Waves and Cosmology, West Virginia University, Chestnut Ridge Research Building, Morgantown, WV 26505, USA}

\author[0000-0002-8509-5947]{Benetge B. P. Perera}
\affiliation{Arecibo Observatory, University of Central Florida, HC3 Box 53995, Arecibo PR 00612, USA}

\begin{abstract}
We report on the latest results of a Parkes multibeam survey for pulsars and dispersed radio bursts in the Large Magellanic Cloud (LMC). 
We conducted both periodicity and single-pulse searches at a much larger range of trial dispersion measures (DMs) than previously searched. We detected 229 single pulses with signal-to-noise ratio ($\rm{S/N}) > 7$ that were classified by the deep learning network FETCH as being real (with $> 90$\% likelihood), of which 9 were from the known giant-pulse-emitting pulsar PSR~B0540$-$69. Two possibly repeating sources were detected with DMs suggesting that they lie within the LMC, but these  require confirmation. Only 3 of the 220 unknown pulses had S/N greater than 8, and the DM distribution for these detected pulses follows an exponential falloff with increasing DM and does not show any excess of signals at DM values expected for the LMC. These features suggest that the detected pulses are not likely to be real, although they are visually compelling. 
We also report the discovery of a new pulsar (PSR J0556$-$67) in our periodicity search. This pulsar has a spin period of 791~ms, a DM of 71~cm$^{-3}$~pc, an estimated 1400~MHz flux density of $\sim 0.12$~mJy, and shows no evidence of binary motion. 
Future observations may be able to confirm whether any of the weak but promising candidates in our single pulse and periodicity searches, including our two possible repeaters, are real or not.
\end{abstract}


\section{Introduction}

Of the more than 3000 pulsars listed in the ATNF pulsar catalog \citep{2005AJ....129.1993M},\footnote{\url{http://www.atnf.csiro.au/research/pulsar/psrcat}} the vast majority ($\sim 99$\%) reside in our Galaxy or its globular clusters. The remainder lies in the Magellanic clouds (MCs) -- Large and Small Magellanic Clouds (LMC and SMC) -- the nearest satellite galaxies to our own. 
There are scientifically motivated reasons for finding and studying pulsars in the LMC.
The discovery of a significant number of radio pulsars in the LMC would provide information about the pulsar population differences between the LMC (an irregular galaxy) and Milky Way (a spiral galaxy) as well as star formation and environmental differences (e.g., metallicity). The LMC has a lower stellar metallicity than the Milky Way (e.g., \citealt{has95}), but the star formation rate in the LMC is estimated to be two orders of magnitude higher than that of the Milky Way per unit mass, which yields a comparable total star formation rate when normalized for galaxy mass \citep{2003MNRAS.339..793G}. Thus, stars in the LMC are clustered despite the comparable star formation rate, which can lead to a higher stellar encounter rate and more massive stars in the LMC than the Milky Way. This is likely to result in the much larger pulsar production rate in the LMC than what the Galactic model would predict.

Pulsar surveys of the LMC may also discover new members of different neutron star (NS) populations. Example source classes include young pulsars, rotating radio transients (RRATs -- rotation powered pulsars that have sporadic bursting emission; \citealt{2006Natur.439..817M}), fast radio bursts (FRBs -- millisecond-duration energetic pulses that have large DMs; \citealt{2007Sci...318..777L}), and millisecond pulsars (MSPs -- old NSs that have been spun up to millisecond periods through a recycling process in binary systems). 
The LMC is a fertile target for finding young NSs, given the large number of supernova remnants per unit mass relative to our Galaxy \citep{2010MNRAS.407.1301B}. This suggests that young NSs manifesting as radio pulsars should be present at a greater rate than in the Galaxy. In fact, two young, rapidly spinning rotation-powered pulsars are known in the LMC: PSR B0540$-$69 \citep{1984ApJ...287L..19S, 1993ApJ...403L..29M} and PSR J0537$-$6910 \citep{1998ApJ...499L.179M}, both of which are X-ray emitters. 

The large number of high-mass X-ray binaries (HMXBs) present in the LMC per unit mass relative to the Galaxy \citep{2005A&A...442.1135L} suggests that there ought to be a higher rate of occurrence of both MSPs and binary systems with unevolved companions (e.g., PSR J0045$-$7319 in the SMC, which orbits a massive, unevolved B star; \citealt{1994ApJ...423L..43K}). Finding and studying binary pulsars with non-degenerate companions may help determine the effect of low metallicity on stellar winds from such companions. These winds can be probed by measuring dispersion measure (DM) and rotation measure variations and depolarization effects induced by the magnetized stellar wind over the course of the orbit (cf.~\citealt{1995MNRAS.275..381M} and \citealt{1996MNRAS.279.1026J} for the PSR~B1259$-$63 system).  White dwarfs (WDs) should also have a lower metallicity in the LMC, and WD pulsar companions may have different properties owing to this. One question that could be answered with a complete sample of binary radio pulsar systems in the LMC (both MSPs and partially recycled pulsars with unevolved companions) is whether the actual number of such systems is consistent with predicted numbers. Binary pulsars are also fossil records of previous binary interactions, and the companion winds and magnetospheres can be studied through the effects of pulsar radiation on them (e.g., \citealt{1996ApJ...459..717K};   \citealt{2009Sci...324.1411A};  \citealt{2005ApJ...634.1223L}).  

In addition to pulsars, our survey probes the LMC for fast transients (including giant radio pulses from pulsars) and holds the possibility for serendipitous discoveries (including FRBs). 
Since only a few members of the underlying RRAT population are visible at any given time owing to their extremely small duty cycles, the number of RRATs present in the Galaxy implied by statistical arguments is large \citep{2006Natur.439..817M}. 
There are currently two known pulsars in the LMC observed to emit giant radio pulses (defined as pulses with flux densities greater than 10 times the mean flux density of the pulsar): PSR B0540$-$69 \citep{2003ApJ...590L..95J, gsa+21} and PSR J0529$-$6652 \citep{2013ApJ...762...97C}. The discovery of more giant pulse emitting pulsars, RRATs, or FRBs in the LMC could play an important role in understanding the populations of these transient sources, which are largely unexplored in the LMC. 
 




To date, no radio MSPs have been discovered in the LMC or anywhere else outside of our Galaxy. MSPs are among the most interesting pulsars to study scientifically, in part because they are generally very stable rotators and are useful ``clocks'' for high-precision timing applications such as the search for gravitational waves \citep{2020ApJ...905L..34A} and tests of gravity \citep{2021PhRvX..11d1050K}. The discovery of stable MSPs in the LMC may impact the search for nanohertz-scale frequency gravitational waves using pulsar timing arrays (PTAs) (\citealt{2015arXiv151107869B}). For example, as shown by \cite{2008ApJ...685.1304L}, the sensitivity of pulsar timing arrays (PTAs) to longitudinal modes of gravitational waves (which are not predicted by general relativity) is directly proportional to the most distant pulsar in the array. Having several baselines with distances of $\sim 50$ kpc, the distance to the LMC  \citep{2013Natur.495...76P}, compared to the rest of a pulsar array at Galactic distances 
could allow such modes to be detected, assuming sufficient pulsar timing accuracy could be achieved.

Pulsar surveys of the MCs also probe the high end of the pulsar luminosity function.
The well-known distance to the LMC provides more accurate luminosity estimates for pulsars discovered there compared to the Milky Way, where pulsar distances can be much more uncertain. Global differences between the luminous pulsar population in the LMC and the Galactic pulsar population may provide insights into the pulsar emission mechanism, such as whether certain emission features might cause radio pulsars to be very luminous. 

In this paper we present the results of a  continuation of a survey for pulsars and transients in the LMC using the Parkes ``Murriyang'' 64-m radio telescope. \citet{2013MNRAS.433..138R} completed  the first 27\% of the planned survey area as defined by the previously most complete pulsar LMC survey conducted with Parkes (\citealt{2006ApJ...649..235M}). The results reported here bring our survey coverage to 40\%. We present the results of a new periodicity search of all of the acquired data to date as well as a new single pulse search these data. 
The paper is organized as follows. In section 2 we describe previous pulsar searches of the LMC, and in section 3, we introduce the details of our new survey. In section 4, we describe the data processing. In section 5, we present our results and discussions. Finally, in section 6, we draw our conclusions.    

%
%
%

\section{Overview of the Survey}

The first Parkes observations of the LMC with the 13-beam multibeam receiver (\citealt{1996PASA...13..243S}) were conducted in 2000 and 2001\footnote{This survey also included observations of the SMC.}, and 136 separate pointings (totaling 1768 beams) covering the LMC were observed (see Fig. 1 of \citealt{2006ApJ...649..235M}). \citet{2013MNRAS.433..138R} also subsequently reprocessed the archival data from these observations of the LMC and discovered five pulsars.

Our Parkes multibeam survey of the LMC began in 2009, in which a new high-resolution backend, the Berkeley–Parkes–Swinburne data recorder (BPSR) was used (\citealt{2010MNRAS.409..619K}). 
The survey plan was to cover the same pattern of 136 interleaved multibeam pointings as in the prior \citet{2006ApJ...649..235M} survey but with better time and frequency channel resolution. This  survey used a center frequency of 1382~MHz and a total bandwidth of 400~MHz split into 1024 channels. The receiver was not sensitive to the top 60~MHz of this band, giving an effective bandwidth of 340~MHz. Similar to the previous multibeam survey of \citet{2006ApJ...649..235M}, which used 8400~s integrations, each pointing was observed for 8600~s, giving a slightly better overall sensitivity from the larger integration time and larger effective bandwidth used (340~MHz vs~288~MHz). A more significant improvement was the increase in sensitivity to fast-spinning pulsars from the shorter sampling time (64~$\mu$s vs~1~ms) and the narrower frequency channel widths (0.4~MHz vs~3~MHz) used. Fig.~2 of \citet{2013MNRAS.433..138R} shows the sensitivity curves for these two surveys for comparison. Our survey is much more sensitive to millisecond pulsars (MSPs), none of which have yet been discovered in the MCs. 

\begin{figure}
\centering
\includegraphics[width=7in]{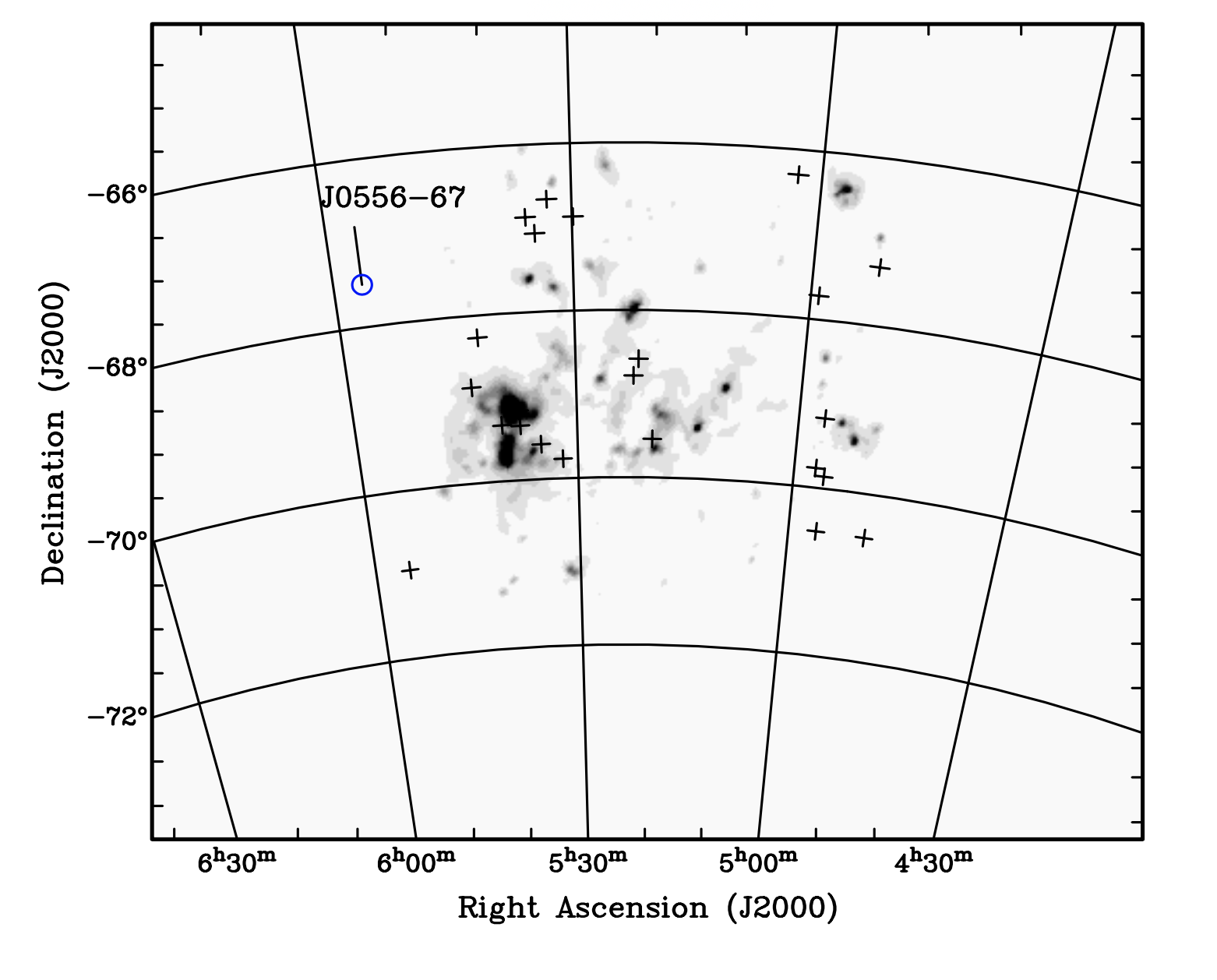}
\caption{The 23 known rotation-powered pulsars in the LMC to date (crosses) plus the newly discovered pulsar PSR J0556$-$67 from our survey analysis (blue circle). The grayscale emission in this figure and subsequent figures is from an IRAS 60 $\mu$m map of the region \citep{2005ApJS..157..302M}. Note that the X-ray pulsar PSR J0537$-$6910 is obscured by the grayscale emission in the 30 Doradus region.}
\label{lmc_pulsars}
\end{figure}

\begin{figure}
\centering
\includegraphics[width=7in]{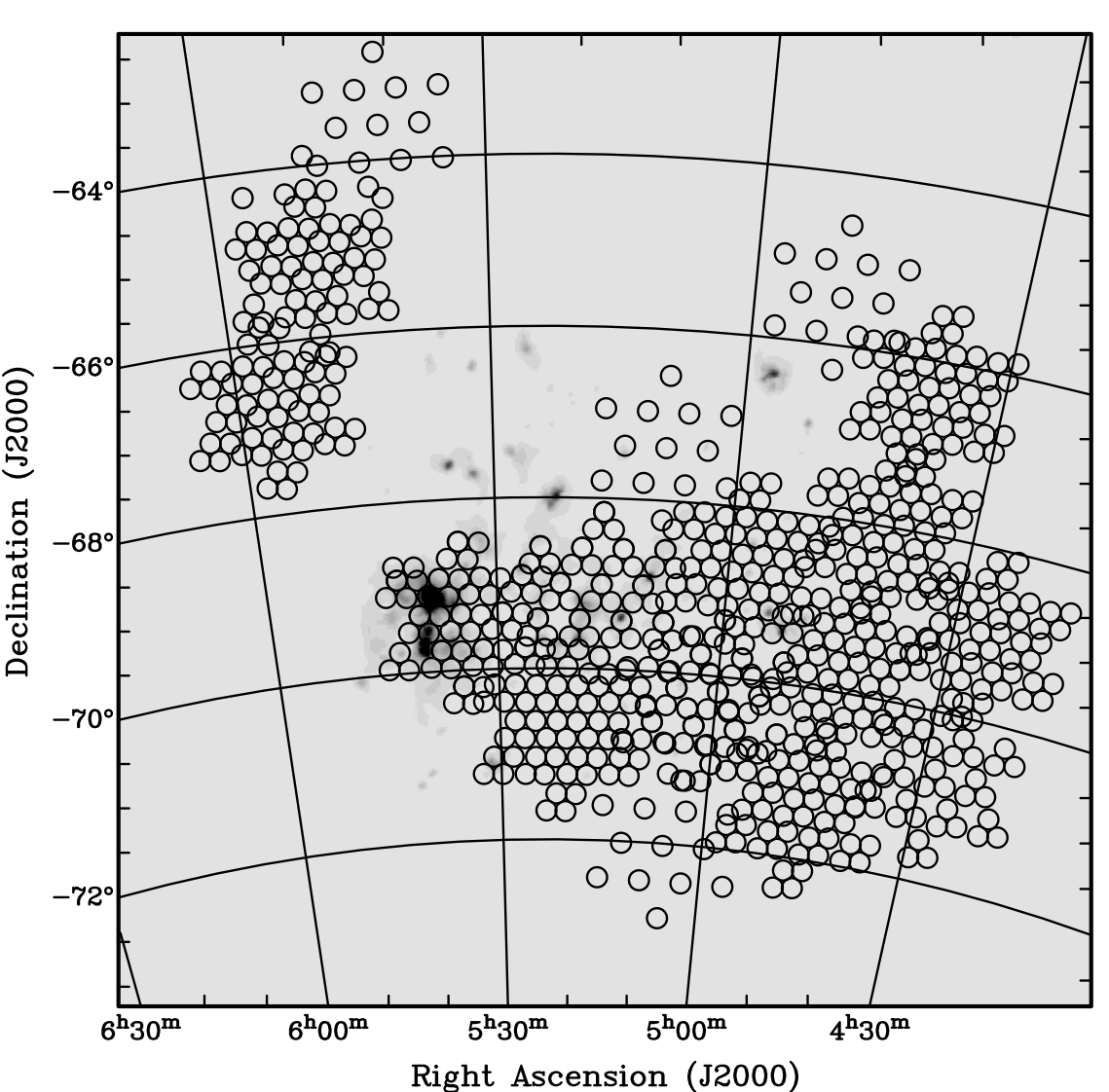}
\caption{Beam coverage of the portion of the survey observed and processed to date. This consists of 54 of the planned 136 pointings (39.7\%) of the full LMC survey coverage of \citet{2006ApJ...649..235M} (see their Fig. 1 therein). Note that each pointing has 13 beams.}
\label{beam_coverage}
\end{figure}

\citet{2013MNRAS.433..138R} discovered three new pulsars in the LMC within the initial 27.2\% of the survey that had been observed as of 2010 (37 unique pointings, plus 3 repeat pointings, corresponding to 520 beams, of which 481 were unique beam positions). This brought the total number of known rotation powered pulsars in the LMC to 23 (see, e.g., \citealt{jpm+22} for a summary). Fig. \ref{lmc_pulsars} shows the locations of these pulsars in the LMC. Subsequent survey data were taken in 2017, which covered an additional 12.5\% of the total survey area,  (17 pointings, corresponding to 221 additional beams). We report on these newer data in this work. Thus, to date 39.7\% of the planned LMC survey pointings have been observed and processed (54 unique pointings out of 136 total planned pointings). Fig.~\ref{beam_coverage} shows this survey coverage of the LMC. 

\section{Data Processing}

We searched for periodic signals with several processing passes through all of the survey data collected to date using different DM search ranges and acceleration trials (the latter to account for binary motion). 
The several passes were done so that we could conduct independent searches with two different software packages and also explore different regions of parameters space in each search.
In the first processing pass, we used \textsc{sigproc}\footnote{\url{http://sigproc.sourceforge.net}} (\citealt{2011ascl.soft07016L}), a widely-used package for pulsar data analysis (see also \citealt{2003ApJ...596.1142C} and \citealt{2016MNRAS.455.2207R}, for example). We searched for signals out to a maximum DM of 800~cm$^{-3}$~pc with an effective sampling time of 128 $\mu$s (pairs of time samples were combined, reducing the time resolution by factor of 2). No acceleration search was performed in this processing pass.
The second pass with \textsc{sigproc} searched DMs between 70 and 300 cm$^{-3}$~pc, a range expected for most LMC pulsars given the DM distribution of the pulsars discovered to date. Only the first 1/8 of each integration was searched, which reduced the raw sensitivity by about a factor of 3.  In this pass, we again combined pairs of samples to give a 128 $\mu$s effective time resolution, and we conducted an acceleration search with acceleration trials in the range of $\pm 6$~m~s$^{-2}$.
The reduced data length and limited range in DM and acceleration we searched in this pass were needed in order to make the processing tractable.
Finally, for the third pass, we used the \textsc{presto}\footnote{\url{https://www.cv.nrao.edu/~sransom/presto}} search and analysis package (\citealt{r01,rem02,2011ascl.soft07017R}). The data were first inspected for radio-frequency-interference (RFI) using the rfifind package, and corrupted channels and time samples were masked (this typically amounted a few percent or less of each observation, with the worst case being 13\%). The time samples were combined into pairs (as above), giving an effective sampling time of 128~$\mu$s. The data were then dedispersed at 5404 DM trials (with variable spacing) in the DM range 0 to 3000 cm$^{-3}$~pc. Each dedispersed time series was Fourier transformed, and the power spectrum was harmonically summed with up to 8 harmonics. Candidate signals were then dedispersed and folded with DMs and periods near the candidate values to look for clear pulsar signals. In this search no acceleration trials were performed.

\citet{2016MNRAS.460.3370C} conducted a single pulse search of the 40 pointings from the LMC survey taken as of 2010 (which includes 3 repeat pointings).\footnote{Note that this set of 40 pointings (520 beams) processed for single pulses is quoted as being 20\% of the survey by \citet{2016MNRAS.460.3370C}. However, this total set of pointings in the percentage calculation includes the full set of 209 LMC and SMC pointigs covered by \citet{2006ApJ...649..235M}, not just the 136 LMC pointings, and also includes the three repeat pointings. Including only the 37 unique pointings as a fraction of the 136 LMC survey pointings gives 27.2\%, which is the same fraction of the LMC observed and processed by \citet{2013MNRAS.433..138R}.}
This was done with \textsc{sigproc}, and the data were searched out to a DM of 5000~cm$^{-3}$~pc. No new signals were found in this analysis. However, \textsc{sigproc} is not optimized for single-pulse detection \citep{2015MNRAS.447.2852K,2019MNRAS.484L.147Z}, and weaker pulses may have been missed in this processing \citep{2019RNAAS...3...41K}. We therefore reprocessed these earlier data and the new data taken in 2017 with a new pipeline based on the \textsc{heimdall}\footnote{\url{https://sourceforge.net/projects/heimdall-astro}} single-pulse search code \citep{b12} in which the candidates are passed through the deep-learning filters developed as part of the \textsc{fetch}\footnote{\url{https://github.com/devanshkv/fetch}} \citep{2020MNRAS.497.1661A} package.

Our single-pulse data processing was divided into two phases. First, we used \textsc{heimdall} to search transient signals at DMs between 50 and 10000~cm$^{-3}$~pc. In the direction of LMC, the maximum Galactic DM contribution is predicted to be $\sim 50$~cm$^{-3}$~pc in both the NE2001 (\citealt{2002astro.ph..7156C}) and YMW16 models (\citealt{2017ApJ...835...29Y}). Hence, the minimum DM in our single pulse search was chosen to be 50~cm$^{-3}$~pc in order to avoid Galactic sources as well as to minimize any RFI effects, which are generally stronger at smaller DMs.
\textsc{heimdall} produced single pulse candidates for each DM trial using boxcar matched filter widths up to 512 samples in each de-dispersed time series (corresponding to a maximum width of 33~ms with our 64~$\mu$s sampling), and we retained candidates with a S/N of at least 7. These candidates were then passed to \textsc{fetch}, a neural network classifier developed for the classification of single pulse candidates to distinguish between RFI and astrophysical signals. RFI can produce a large number of candidates that have similar characteristics to astrophysical signals. \textsc{fetch} assigned a probability to each detected candidate of being a real, astrophysical signal, and we selected candidates for consideration which had a \textsc{fetch}-assigned probability of at least 90\%. As a further check, each of these selected candidates was also individually inspected visually to ensure that there were no obvious RFI signals that were incorrectly classified by \textsc{fetch} as real.

\section{Results and Discussion}

\begin{figure}
\centering
\includegraphics[width=3.5in]{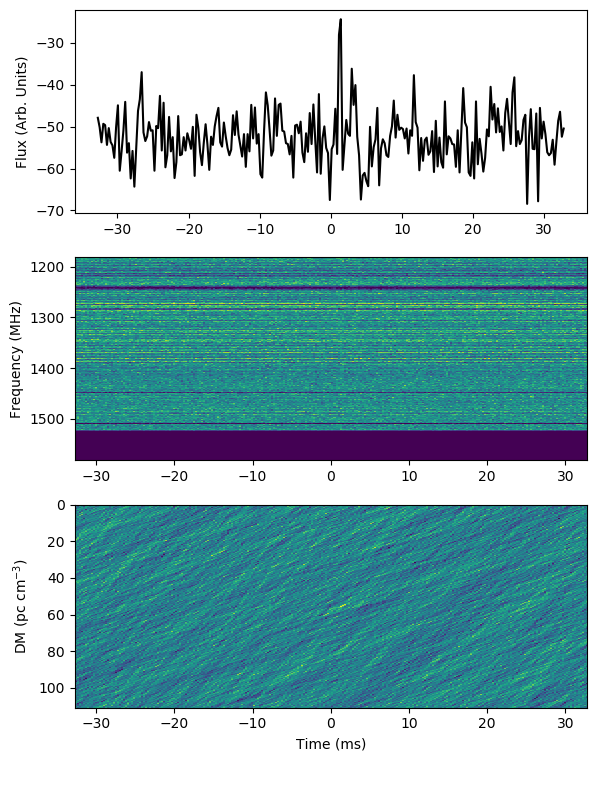}
\includegraphics[width=3.5in]{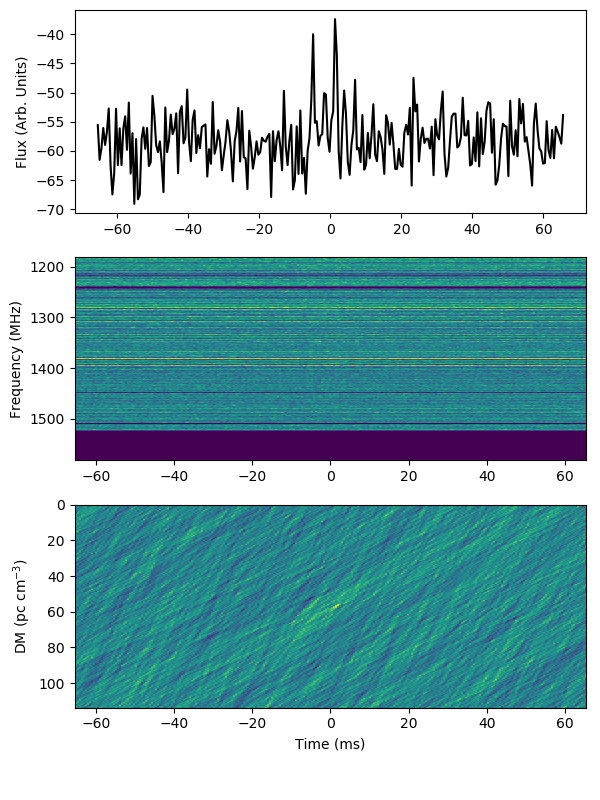}
\caption{A possibly repeating signal at a DM of 56~cm$^{-3}$~pc (see Table \ref{table:list}).}
\label{repeater1}
\end{figure}

\begin{figure}
\centering
\includegraphics[width=3.5in]{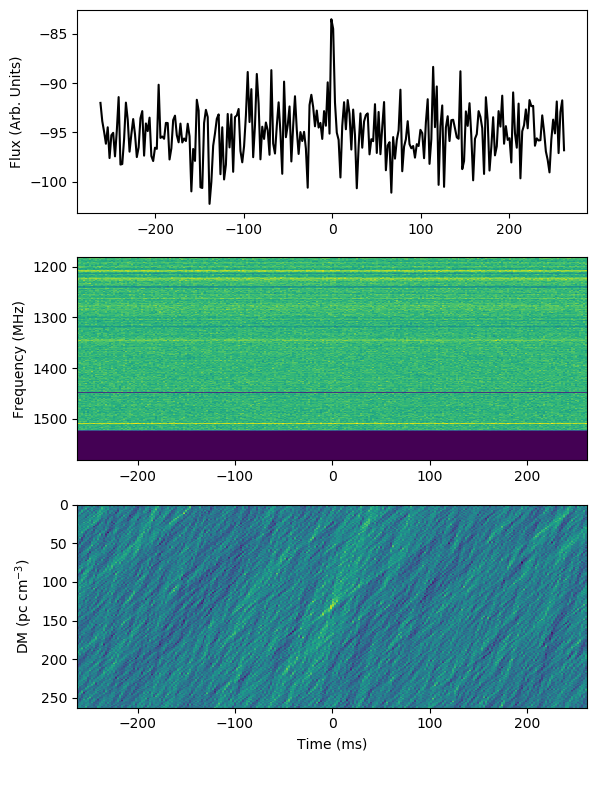}
\includegraphics[width=3.5in]{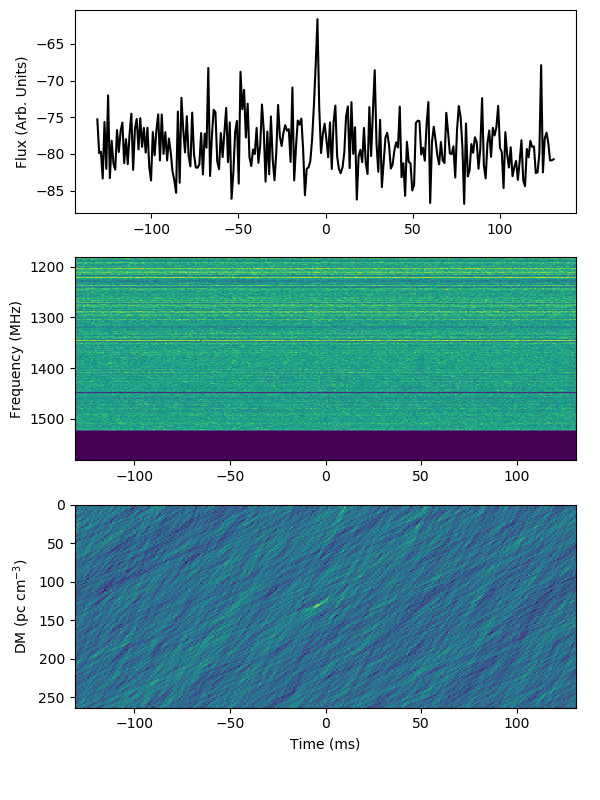}
\caption{A possibly repeating signal at a DM of 131~cm$^{-3}$~pc (see Table \ref{table:list}).}
\label{repeater2}
\end{figure}

We detected a total of 229 single-pulse candidates in 187 survey beams that met our selection criteria. 
The number of candidates produced by HEIMDALL for the observed survey beams is $6.3 \times 10^{5}$. This is much greater than the number of FETCH candidates which passed our criteria with a probability of 0.9, which is 229, including 9 pulses from B0549-69. Thus, the classifier is useful in reducing this huge number of candidates. Two of these beams had two detected signals each which occurred at the same DM (56 and 131 cm$^{-3}$~pc, respectively), suggesting that they could be repeating sources (see Figs.~\ref{repeater1} and \ref{repeater2}). Both of the DMs of these possible repeaters are consistent with the DM range seen for other pulsars in the LMC (see Table 2 of \citealt{2013MNRAS.433..138R}), indicating that if they are real, then they are originating from the LMC and not the foreground. We searched these two beams for accelerated pulsars using \textsc{PRESTO} periodicity searches  at these specific DMs, but we detected no signals.

\begin{figure}
\centering
\includegraphics[width=3.5in]{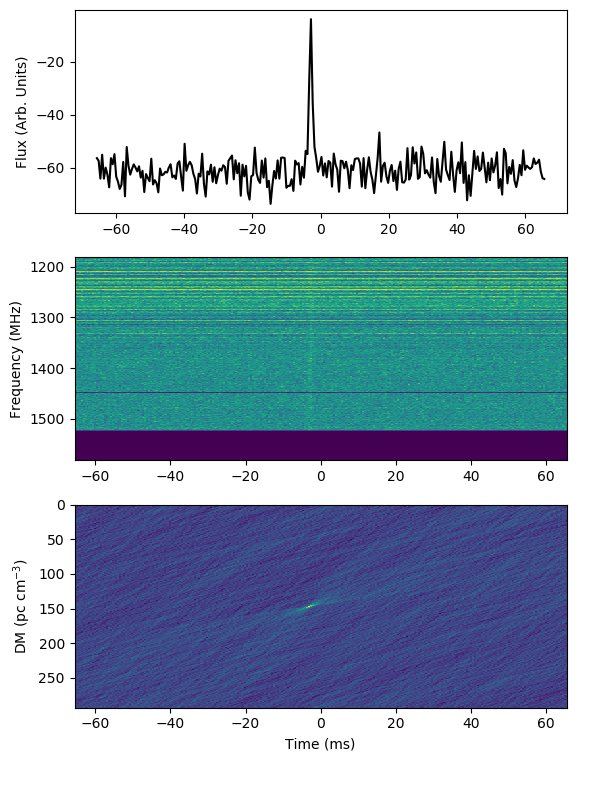}
\caption{The strongest detection (S/N $\sim$ 18) of PSR~B0540$-$69 in our blind single pulse search. This was one of eight single-pulse detections made from this pulsar.}
\label{B0540-69}
\end{figure}

We also clearly detected 9 pulses from the known giant pulse emitting pulsar PSR~B0540$-$69 \citep{2003ApJ...590L..95J}. These were among the strongest detections in our sample and were clearly identified by \textsc{fetch} as being real with high probabilities. Fig.~\ref{B0540-69} shows the strongest of these 9 detections. Of the 10 total single-pulse candidates with S/N $>$ 7 from B0549-69 found by HEIMDALL, 9 were identified by FETCH with probability $>$ 0.9, which is our threshold for consideration as real. The remaining pulse had a probability of 0.8. FETCH was able to reliably identify almost all of the pulses above S/N of 7. No survey beams were coincident with PSR J0529$-$6652, the other LMC pulsar known to be detectable in single pulses \citep{ 2013ApJ...762...97C} (Fig. \ref{beam_coverage}). The properties of the pulses from these two possible repeating sources and the pulses detected from PSR~B0540$-$69 are listed in Table \ref{table:list}. 

There were 220 unidentified pulses that remained after excluding the 9 pulses detected from PSR~B0540$-$69. \citet{2021MNRAS.507.3238Y} reported 81 new FRB candidates discovered by reprocessing archival data from the Parkes multibeam survey. None of their candidates are spatially coincident with the LMC, which could have provided possible confirmations of some of our candidates through repeat detections.

\begin{figure}
\centering
\includegraphics[width=7in]{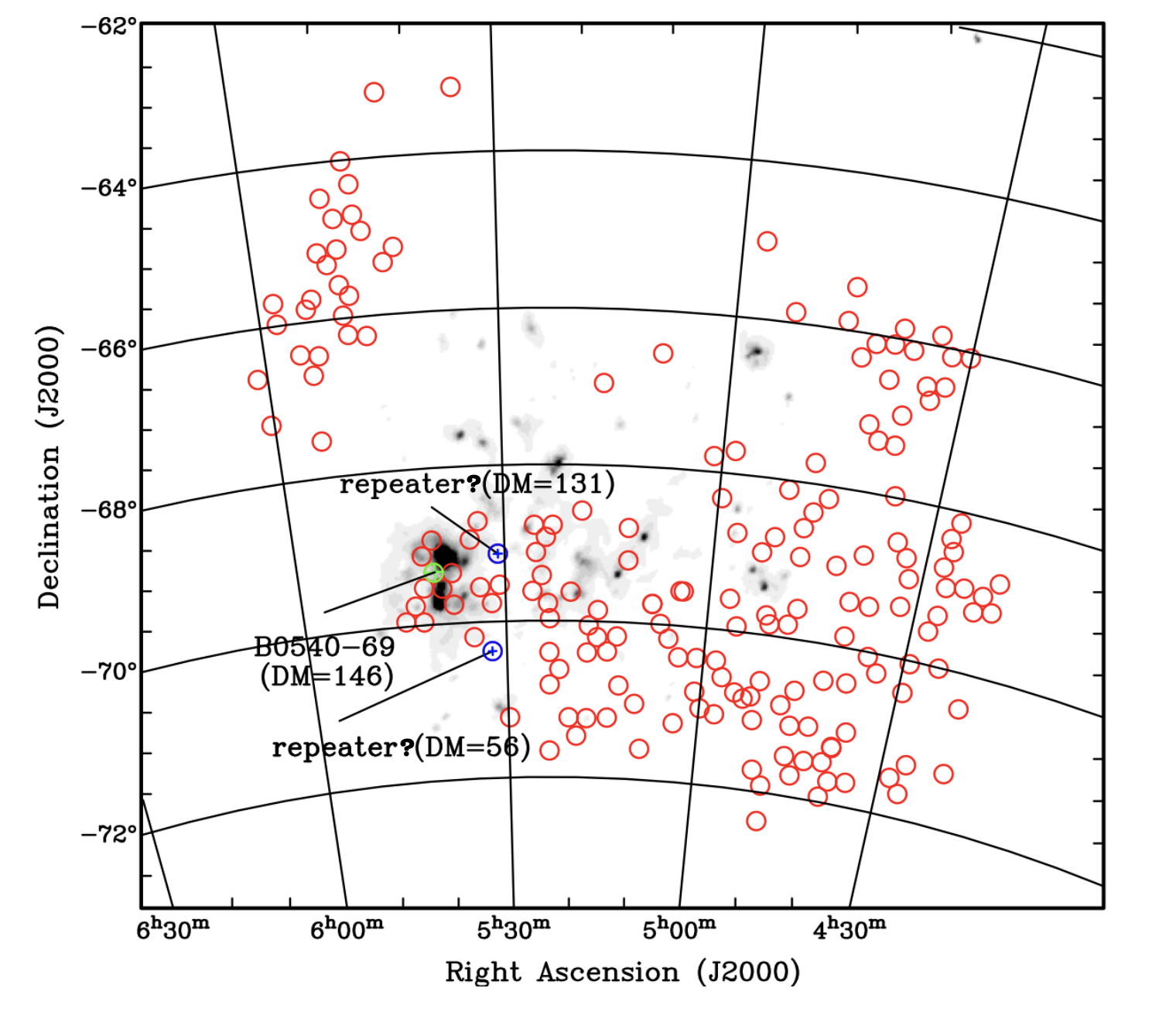}
\caption{The 187 survey beams with at least one single-pulse candidate detected and identified by FETCH with a probability of being real that exceeds 90\% (red circles). This represents more than 25\% of the total number of beams that have been observed and processed. The blue circles with crosses indicate the two beams where two signals were detected in a single integration at the same DM, suggesting possible repeating sources. The DMs of these signals (56 and 131~cm$^{-3}$~pc) are indicated on the plot. The green circle indicates the beam in which multiple giant pulses from PSR~B0540$-$69  were clearly detected. The position of PSR~B0540$-$69 is marked by the green cross.}
\label{fig-sps}
\end{figure}

Fig.~\ref{fig-sps} shows a map of the 187 beams where at least one single-pulse candidate was identified by \textsc{fetch} with at least a 90\% probability of being real. Note that this represents more than 25\% of the total number of beams in the survey observed and processed to date. The green and blue circles in Fig. \ref{fig-sps} indicate the positions of PSR~B0540--69 and the two possible repeater candidates, respectively. Like PSR~B0540--69, the possible repeater candidates also reside in the 30 Doradus star forming region. 

\begin{figure}
\centering
\includegraphics[width=7.5in]{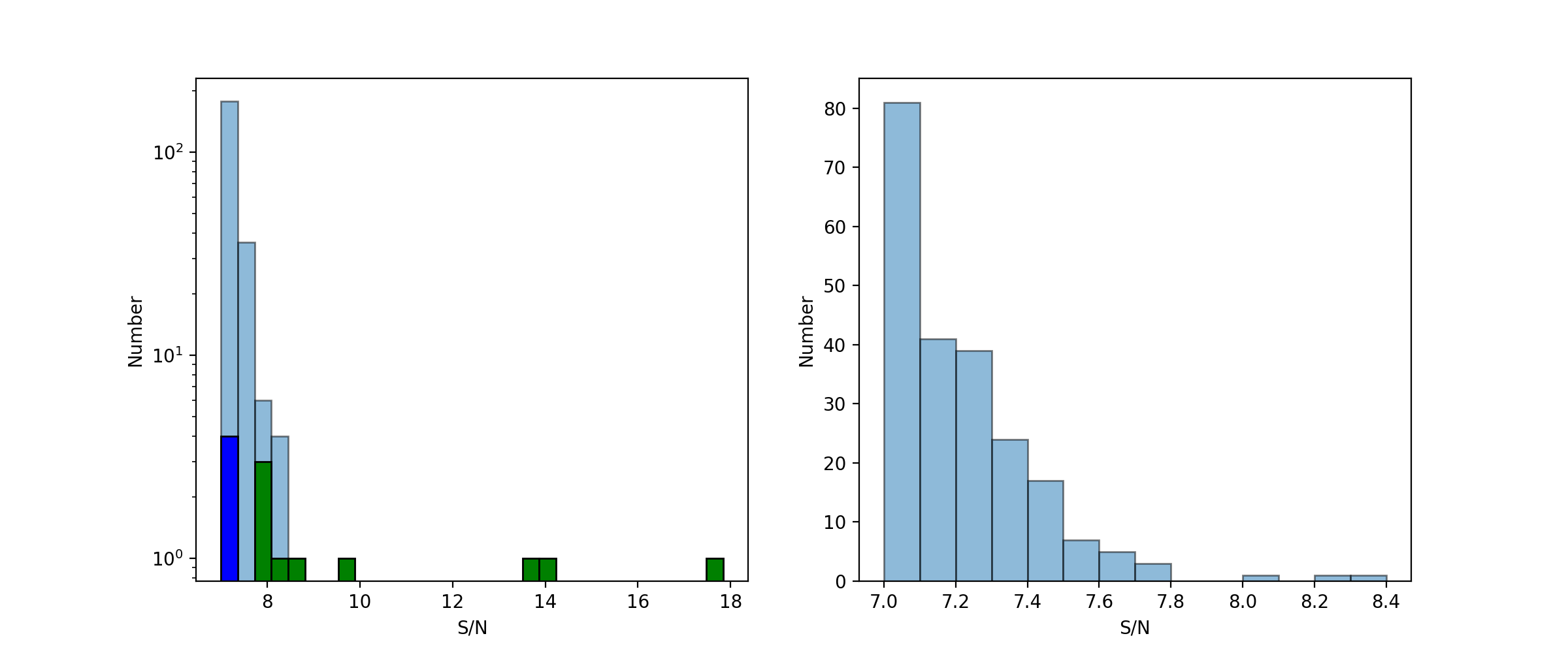}
\caption{(Left) Distribution of S/N for 229 detected single pulses identified by \textsc{fetch} with high confidence ($> 90$\%) of being real and with a realistic morphology. The dark blue bar contains the 4 signals from the two possibly repeating sources. The green bars indicate the 9 single pulse detections from the known giant pulse emitter PSR~B0540$-$69. (Right) S/N distribution of all 220 unidentified signals (i.e., excluding the 9 pulses from PSR~B0540$-$69). The vast majority of these signals 
are weak, with $\rm{S/N} < 8$.}
\label{SN_hist}
\end{figure}

A histogram of the S/N distribution of all 229 detected pulses is shown in Fig. \ref{SN_hist} (left). The candidates from the possible repeaters (dark blue bars) and from PSR~B0540$-$69 (green bars) are included. Excluding the 9 known pulses from PSR~B0540$-$69 leaves all but 3 of the 220 remaining candidates (98.6\%) with a S/N below 8, including the two possible repeating sources. 
This is illustrated further in Fig.~\ref{SN_hist} (right). The cutoff found in the histogram above about 7.8 indicates that there are no strong detections other than giant pulses from PSR~B0540--69. This calls into question whether many of these weaker pulses are in fact astrophysical in origin, despite their promising morphologies.

\begin{figure}
\centering
\includegraphics[width=7.5in]{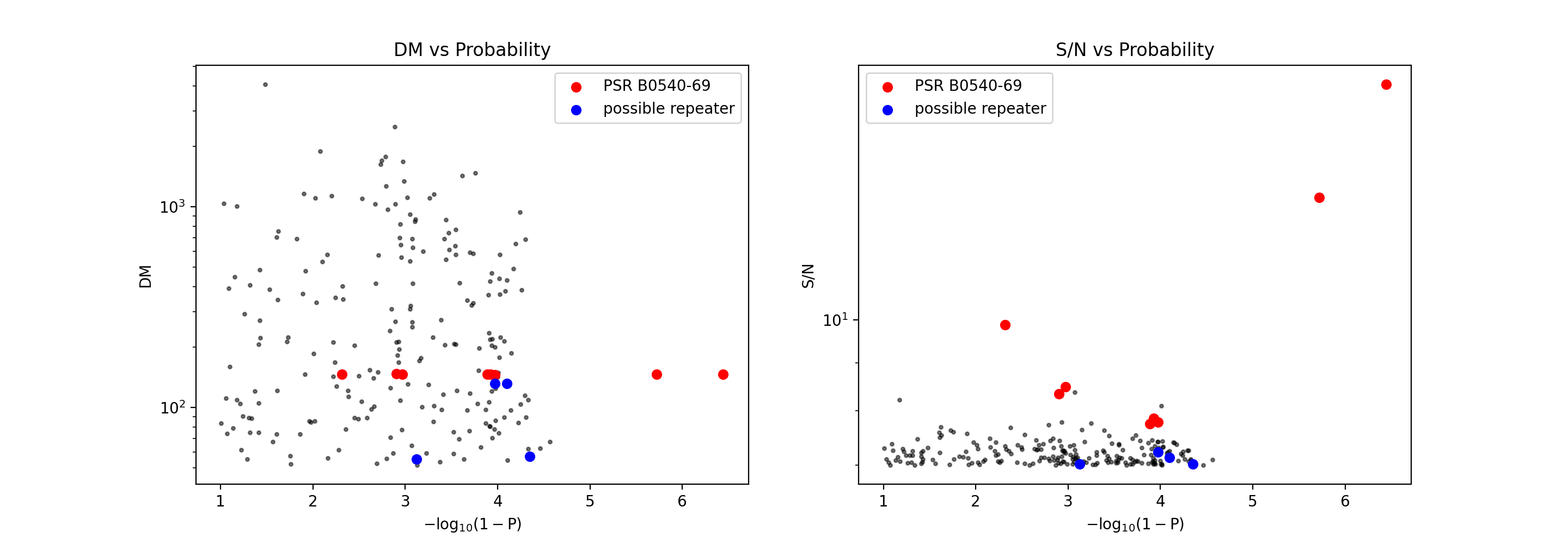}
\caption{Single pulse events identified by FETCH with high confidence ($> 90$\%) of being real and with a realistic morphology. The vertical axis is DM (left) and S/N (right) and the horizontal axis is $-\log_{10}(1-P)$, where $P$ is the FETCH probability of being real (points in the right of the plot have been identified as more likely to be real). In both panels, red points indicate the 8 pulses detected from PSR~B0540$-$69, and blue points indicate the possibly repeating signals (two sets of two pulses). Note that one single pulse from PSR B0540-69 is excluded because its probability is 1 (i.e., $-\log_{10}(1-P)$ is infinite)}
\label{relationship}
\end{figure}

Fig.~\ref{relationship} shows DM and the \textsc{fetch}-assigned probability of being real (left) and S/N and probability (right). In these plots, the probability is restated in terms of its logarithm, $-\log(1- P)$, in order to show the range more clearly. Five of the eight signals from PSR~B0540$-$69 and 3 of the 4 signals from the possible repeaters have slightly higher probabilities than bulk of the detections ($-\log(1- P) \ga 4$). However, the signals from PSR~B0540$-$69 in Fig.~\ref{relationship} (right) show that a higher S/N for a real signal does not necessarily lead to a higher \textsc{fetch} probability. The candidates are generally distributed uniformly as a function of the log of the probability, which indicates that there is no clear relationship between DM or S/N and the logarithm of the probability. 
Most of candidates identified by \textsc{fetch} as having a high probability of being real were nevertheless weak (S/N~$< 8$). A visual check of these candidates suggests that they appear real, and \textsc{fetch} also clearly identified the 9 giant pulses from PSR~B0540$-$69 as being real (i.e., with probability exceeding 99\% in all cases; see Table \ref{table:list}), thus confirming the ability of \textsc{fetch} to accurately identify real pulses. 

However, in a survey conducted with Arecibo, \citet{2022MNRAS.509.1929P} searched for single-pulse candidates with \textsc{heimdall} and \textsc{fetch} over negative DM values in order to assess whether spurious, non-astrophysical candidates could be produced which mimicked real candidates.  We conducted a similar single-pulse search with negative DM values ranging between $-$10000 and $-$50 cm$^{-3}$~pc for one our survey pointings, corresponding to 13 beam positions. This processing resulted in a large number of weak, artificial candidates being detected, which exceeded the number of candidates produced in the positive DM search. As in the case of \citet{2022MNRAS.509.1929P}, this suggests that many (if not all) of these low S/N candidates we detected in our search are actually likely to be artificial, non-astrophysical signals.

Histograms showing the DM distribution for the candidates identified by FETCH are shown in Fig. \ref{histogram}, with the panel on the right showing only a subset of signals with DMs between 50 and 400~cm$^{-3}$~pc. This is the DM range expected for pulsars in the LMC given the currently known LMC pulsar population (the largest DM of a known LMC pulsar is 273~cm$^{-3}$~pc; \citealt{2013MNRAS.433..138R}). The error bars were computed as the square root of the number of pulses in each bin, and the number of bins $k$ was determined according to the Sturges criterion, $k = 1 + \log_{2} n$, where $n$ is the number of pulse events detected. Pulses from PSR~B0540$-$69 were excluded from these histograms. The DMs of about half of the candidates we detected are larger than the highest DM in the known LMC pulsar population (273 cm$^{-3}$~pc), and some single pulse candidates we identified had DM $> 1000$ cm$^{-3}$~pc, which is well above this range. If these signals are real, this suggests that they either come from the most distant objects in the LMC, or, more likely, that they are not real. Both DM distributions (the full sample and the subset) follow an exponential falloff as a function of DM (Fig.~\ref{histogram}), $N(DM) = A\exp(-DM/B)$, though not with the same fit parameters in each case. Table \ref{parameters} shows the best-fit parameters with uncertainties.
Integrating this best fit exponential falloff in the histogram for the subset of signals between 50 and 400 cm$^{-3}$~pc gives an expected number of pulses above DM 1000 cm$^{-3}$~pc of about 0.3 (i.e., less than 1). Our observed sample of 22 detections above this range is clearly not consistent with this expected number and supports the interpretation that
these signals are not real. 
In Fig.~\ref{histogram} (right), the sample shows no clear excess of signals in the likely range of DM values for the LMC (i.e., 50 to 400~cm$^{-3}$~pc). This also suggests that most of these signals are not likely to be real.

We estimated the probability of detecting two single pulses by chance in the same beam at the same DM with FETCH probability $>$ 0.9. The fitting equation and its parameters in the DM range of 50 - 400 pc cm$^{-3}$ (see Table \ref{parameters}) give us the expected number of candidates per beam, which are 0.084 and 0.040 for DMs of 56 and 131 pc cm$^{-3}$, respectively. From the Poisson distribution, the  probabilities of finding chance repeat signals at these DMs are 0.32$\%$ and 0.08$\%$, respectively. Thus, these repeat events are not likely to occur by chance. 


In our periodicity search, we have discovered a new pulsar in the LMC, PSR J0556$-$67 (Fig.~\ref{new_pulsar}; see also Fig. \ref{lmc_pulsars} for its position) in both the  \textsc{SIGPROC} and \textsc{PRESTO} pipelines. This 791 ms period pulsar was found in survey data acquired after the publication of \citet{2013MNRAS.433..138R}. Its DM of 71~cm$^{-3}$~pc lies within the range of other pulsars discovered in the LMC. In the discovery observation, there is no significant measured period derivative (and hence no clear evidence of a binary motion), which is consistent with expectations for a canonical long-period pulsar. The position determination of PSR J0556$-$67 is accurate to about 7~arcmin from the nominal discovery position, corresponding to the radius of the Parkes discovery beam. We estimate a 1400~MHz flux density for the pulsar of $\sim 0.12$~mJy (Fig.~\ref{flux&period}) using Eq.~2 in \citet{2013MNRAS.433..138R}, but we do not yet have timing information for this pulsar. No other new pulsars were found in our search, though we did find a number of weak periodicity candidates that could be confirmed in future observations.

\section{Conclusions}

We report on a survey of the LMC for pulsars and fast transients using the Parkes multibeam receiver with high time and frequency resolution that was initiated by \citet{2013MNRAS.433..138R}. To date we have observed and processed  40\% of the total planned LMC survey area defined by the \citet{2006ApJ...649..235M} survey coverage. We have conducted both new periodicity and single-pulse searches of all of the data acquired to date.
We detected 229 single-pulse candidates above a S/N of 7 having a \textsc{fetch}-assigned probability of being real exceeding 90\%. Of these, 9 pulses were from the known giant radio pulse emitting pulsar PSR~B0540$-$69 (see Table~\ref{table:list}), and 220 were from unidentified sources. Two possibly repeating single-pulse sources were detected at DMs of 56 and 131 cm$^{-3}$~pc (see Table~\ref{table:list}), which is in the range expected for the LMC. However, a single-pulse search of negative DM values indicates that realistic, low S/N ($<$ 8) candidates can be artificially produced. We conclude that the majority of our detected candidates are not likely to be real. The two possible repeating sources are also not definitively real, and future observations of these targets will be essential to claim real detections.

\begin{figure}
\centering
\includegraphics[width=7in]{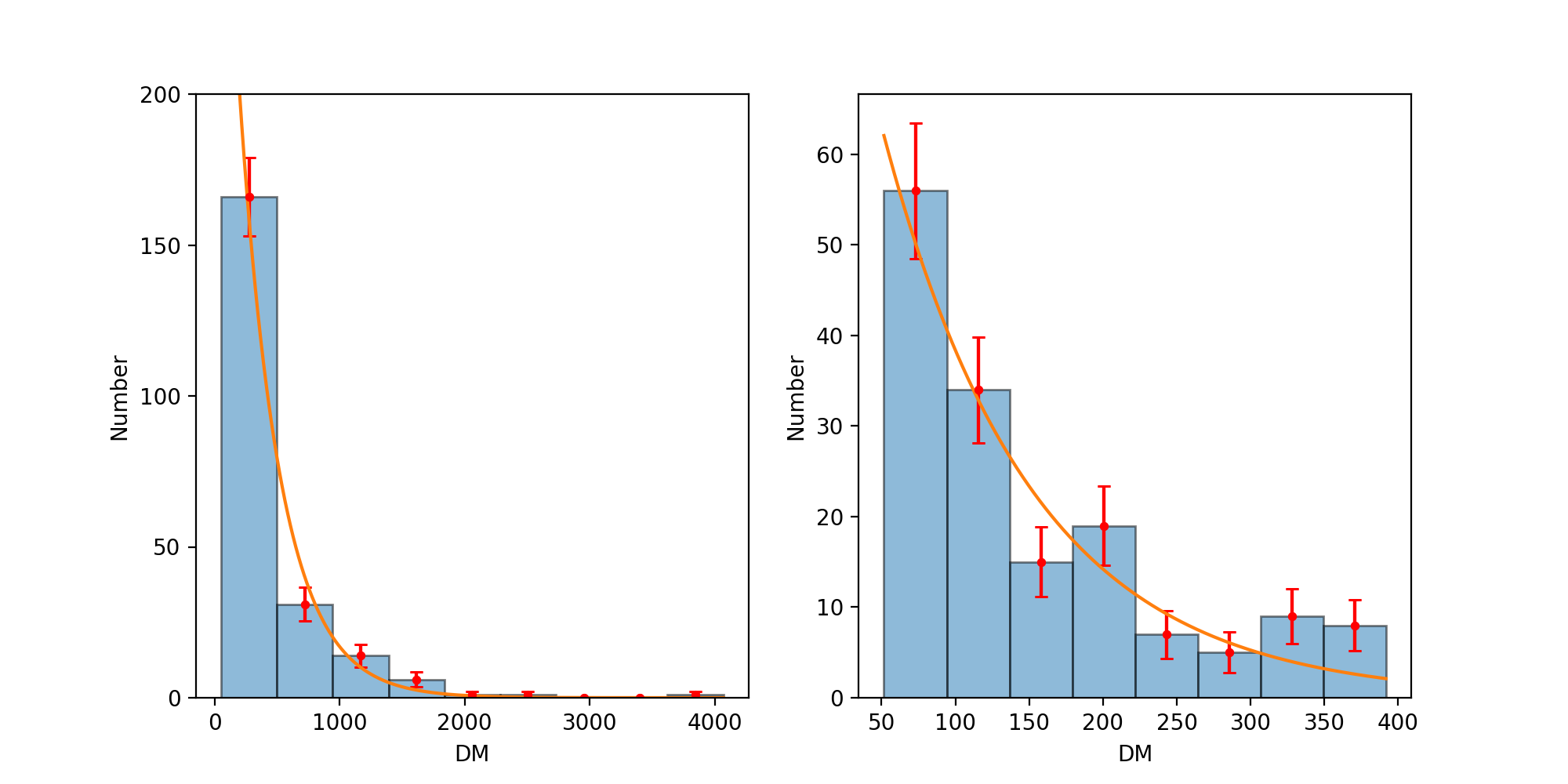}
\caption{DM histogram of the 220 single pulse candidates identified by FETCH with high confidence ($> 90$\%) of being real (excluding the 9 known pulses detected from PSR~B0540$-$69). Both the full DM range of the detected signals (DMs above 50 cm$^{-3}$~pc; left) and a subset (DMs between 50 and 400 cm$^{-3}$~pc; right) are shown. Error bars were computed as the square root of the number of pulses in each bin. In both cases, a weighted exponential fit was performed and is overlaid on the plot (the fit parameters differed in each case). No obvious excess of signals was detected in the DM range expected for LMC pulsars, as determined from the DMs of known LMC pulsars.}
\label{histogram}
\end{figure}

\begin{figure}
\centering
\includegraphics[width=7in]{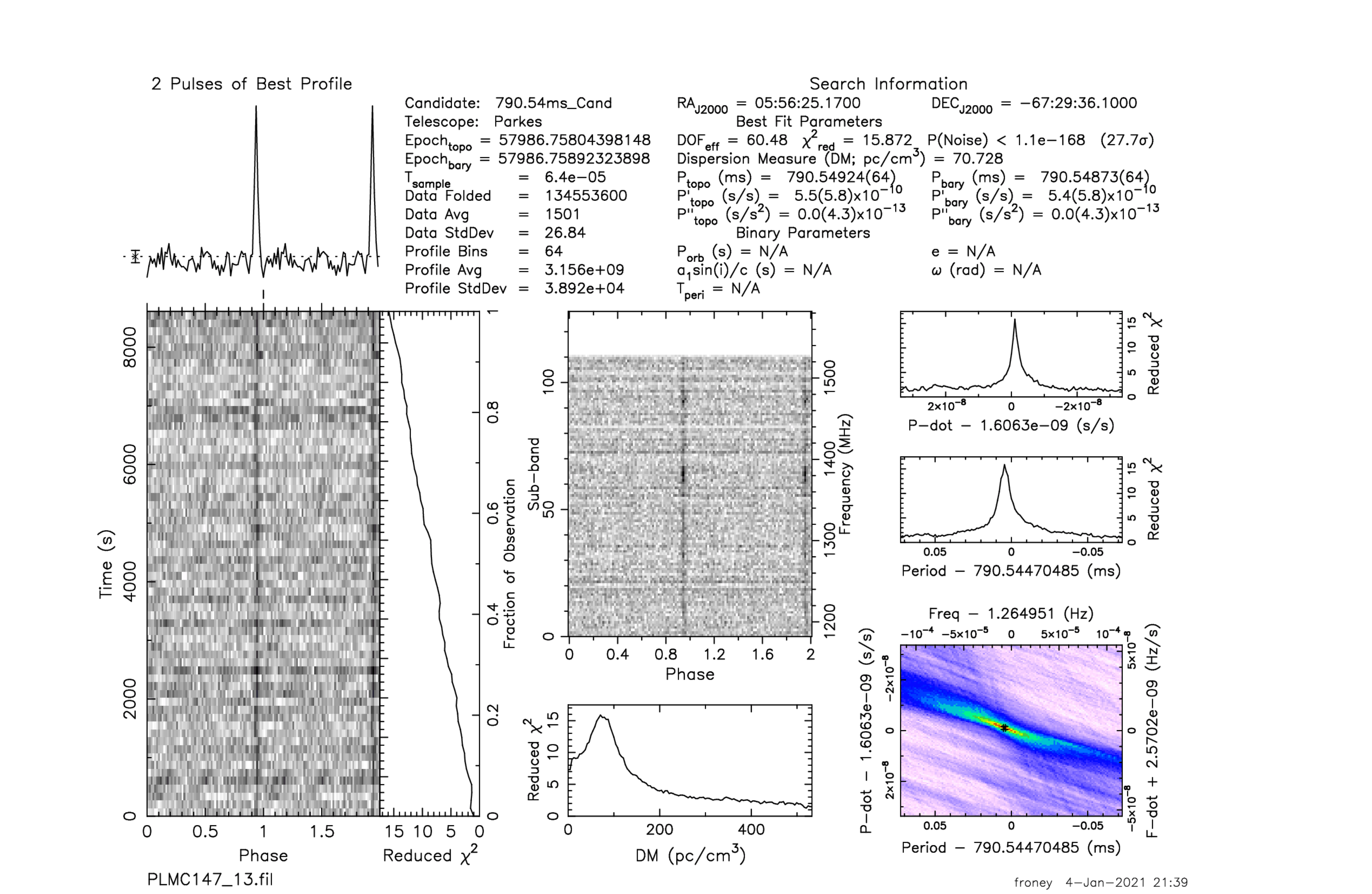}
\caption{\textsc{presto} prepfold plot of PSR J0556$-$67, a 791-ms pulsar discovered in our survey. The DM of 71 cm$^{-3}$~pc indicates that it resides in the LMC. No timing information is yet available. Its 1400~MHz flux density is estimated to be $\sim 0.12$~mJy.}
\label{new_pulsar}
\end{figure}

\begin{figure}
\centering
\includegraphics[width=7in]{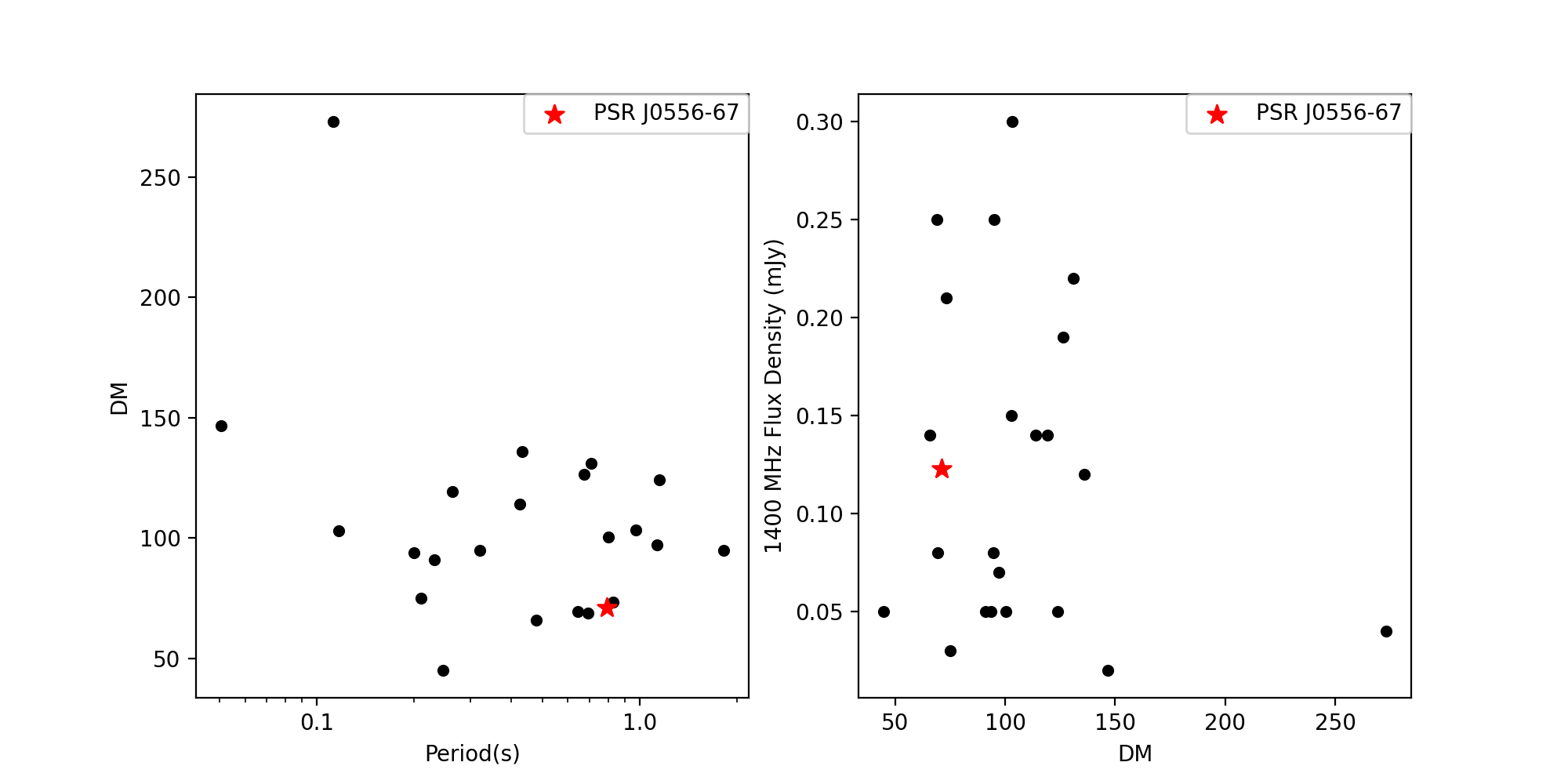}
\caption{DM vs. period for all 23 radio pulsars in the LMC (left), including the new discovery PSR J0556$-$67 (but not including the X-ray pulsar PSR J0537$-$6910). PSR J0556$-$67 is indicated by the star. The flux density vs. DM at 1400~MHz is shown for these same 23 known radio pulsars (right). See also \citet{2013MNRAS.433..138R}.}
\label{flux&period}
\end{figure}

From an exponential fit to the DM histogram of single pulses detected with DMs between 50 and 400~cm$^{-3}$~pc (a likely range for pulsars in the LMC), we would expect to find less than one instance of a signal above a DM of 1000~cm$^{-3}$~pc (see Fig.~\ref{histogram}). This is significantly smaller than the 22 candidates detected above DM = 1000~cm$^{-3}$~pc and also suggests that those signals are likely not real.  These high-DM signals appear no different qualitatively than the many lower-DM signals we found, which lends additional support to the conclusion that all or most of our detected signals are artificial despite their promising morphology.

In our periodicity search, we discovered one new pulsar in the LMC, PSR J0556$-$67, which has a spin period of 791 ms and a DM of 71 cm$^{-3}$~pc. We estimate its flux density to be $\sim 0.12$~mJy. Its DM is consistent with the population of known LMC pulsars, and brings the total number of rotation-powered pulsars in the LMC (including the X-ray pulsar PSR J0537$-$6910) to 24. No timing information is yet available for this pulsar. We have discovered no new MSPs in our survey, and to date no MSPs have been discovered outside of our Galaxy. 

To complete the remaining 60\% of the survey area, we would require an additional 200 hr of observing time with the multibeam system at Parkes. This is unlikely to occur given that the MeerKAT telescope is slated to survey the LMC with better sensitivity than Parkes. However, future targeted survey observations of the 30 Doradus region of the LMC, where both PSRs B0540$-$69 and J0537$-$6910 reside (as well as our two possibly repeating single pulse sources), may be productive given the apparently high concentration of young pulsars there. Future observations with Parkes with its more sensitive ultra-wide-bandwith low-frequency receiver (spanning frequencies between 700~MHz and 4~GHz) (\citealt{2020PASA...37...12H}) or observations from the Transients and Pulsars with MeerKAT (TRAPUM) project\footnote{\url{http://www.trapum.org}} (\citealt{2016mks..confE...9S}; \citealt{2020PASA...37...28B}) may be able to confirm some of our weak periodicity candidates or single-pulse candidates we detected.

\begin{acknowledgments}

The Parkes radio telescope is part of the Australia Telescope National Facility (grid.421683.a) which is funded by the Australian Government for operation as a National Facility managed by CSIRO. We acknowledge the Wiradjuri people as the traditional owners of the Observatory site. SH is supported by JSPS KAKENHI Grant Number 20J20509 and the JSPS Overseas Challenge Program for Young Researchers. KT is partially supported by JSPS KAKENHI Grant Numbers 20H00180, 21H01130, and 21H04467, Bilateral Joint Research Projects of JSPS, and the ISM Cooperative Research Program (2021- ISMCRP-2017).
 
We acknowledge the use of software in this work. 
\software{sigproc (\citealt{2011ascl.soft07016L}), PRESTO (\citealt{r01,rem02,2011ascl.soft07017R}), HEIMDALL \citep{b12}, FETCH \citep{2020MNRAS.497.1661A}, Karma \citep {1996ASPC..101...80G}}
\end{acknowledgments}

\begin{table}
\begin{center}
\caption{Single pulse detections from two possibly repeating sources and from the known pulsar PSR~B0540$-$69. Beam positions have units of right ascension of hours, minutes, and seconds, and units of declination of degrees, arcminutes, and arcseconds. The uncertainty in the position of the possibly repeating sources is about 7~arcmin, corresponding to the Parkes beam radius. The last three columns give the detected DM, S/N, and the \textsc{fetch}-assigned probabilities for these candidates, respectively.} 

\begin{tabular}{lccccc}
\hline
Name & Beam RA (J2000) & Beam Dec (J2000) & DM & S/N & \textsc{fetch} Probability \\ 
     & (hh:mm:ss)      & (dd:mm:ss)       & (cm$^{-3}$~pc)  & \\ \hline
Repeater? & 05:20:30.0 & $-$71:14:09.4 & \multicolumn{1}{c}{55.4} & \multicolumn{1}{c}{7.02} & 0.9992490 \\
     &         &          & \multicolumn{1}{c}{56.9} & \multicolumn{1}{c}{7.02} & 0.9999553 \\ \hline
Repeater? & 05:30:57.8 & $-$69:08:04.0 & \multicolumn{1}{c}{131.8} & \multicolumn{1}{c}{7.13} & 0.9999205   \\
     &         &          &  \multicolumn{1}{c}{131.4}  & \multicolumn{1}{c}{7.23} & 0.9998933 \\ \hline
PSR~B0540$-$69 & 05:40:20.7 & $-$69:20:15.0 & \multicolumn{1}{c}{146.4} & \multicolumn{1}{c}{7.75} & 0.9998703 \\
& & & \multicolumn{1}{c}{145.5} & \multicolumn{1}{c}{7.78} & 0.9998932 \\
& & & \multicolumn{1}{c}{146.4} & \multicolumn{1}{c}{7.86} & 0.9998814 \\
& & & \multicolumn{1}{c}{146.9} & \multicolumn{1}{c}{8.34} & 0.9987482 \\
& & & \multicolumn{1}{c}{146.4} & \multicolumn{1}{c}{8.49} & 0.9989248 \\
& & & \multicolumn{1}{c}{146.4} & \multicolumn{1}{c}{9.88} & 0.9951752 \\
& & & \multicolumn{1}{c}{146.4} & \multicolumn{1}{c}{13.51} & 0.9999981 \\
& & & \multicolumn{1}{c}{146.0} & \multicolumn{1}{c}{13.884} & 1 \\
& & & \multicolumn{1}{c}{146.4} & \multicolumn{1}{c}{17.84} & 0.9999996 \\
\hline
\end{tabular}
\label{table:list}
\end{center}
\end{table}

\begin{table}
\centering
\caption{Best-fit parameters for exponential function fits to the DM histograms shown in Fig.~\ref{histogram}. The fitted parameters were defined according to $N(DM) = A\exp(-DM/B)$. Quoted uncertainties are 1$\sigma$ errors.}
\begin{tabular}{wc{5cm}|Wc{4cm}|Wc{4cm}}
\hline
\multicolumn{1}{c}{DM Range}& \multicolumn{1}{c}{$A$} & \multicolumn{1}{c}{$B$}\\ \hline
\multicolumn{1}{l}{$50 <$ DM $< 10000$~cm$^{-3}$~pc (full DM range)} & \multicolumn{1}{c}{$370 \pm 60$} & \multicolumn{1}{c}{$323 \pm 34$} \\ 
\multicolumn{1}{c}{$50 <$ DM $< 400$~cm$^{-3}$~pc (LMC expected only)} & \multicolumn{1}{c}{$103 \pm 28$} & \multicolumn{1}{c}{$100\ \pm 17$}\\ \hline
\end{tabular}
\label{parameters}
\end{table}


\bibliography{ref}{}
\bibliographystyle{aasjournal}
\end{document}